\documentclass[reprint,amsmath,amssymb,aps,pre,showpacs,eqsecnum]{revtex4-1}

\usepackage{graphicx}
\usepackage{dcolumn}
\usepackage{bm}

 \usepackage{color}

\usepackage{amssymb}
\usepackage{graphicx}
\usepackage{amsbsy}

\newcommand\beq{\begin{equation}}
\newcommand\eeq{\end{equation}}
\newcommand\beqa{\begin{eqnarray}}
\newcommand\eeqa{\end{eqnarray}}
\newcommand{\nn}{\nonumber\\}

\newcommand{\rr}{\mathbf{r}}

\newcommand{\NN}{\mathbf{N}}

\newcommand{\rN}{\rr^N}

\newcommand{\ZZ}{\mathcal{Z}}

\newcommand{\dd}{\text{d}}
\newcommand{\ex}{\text{ex}}
\newcommand{\id}{\text{id}}
\newcommand{\py}{\text{PY-}}

\newcommand{\xx}{{(\xi)}}

\begin{document}



\title{Chemical-potential route for multicomponent fluids}


\author{Andr\'es Santos}
\email{andres@unex.es}
\homepage{http://www.unex.es/eweb/fisteor/andres/}

\affiliation{Departamento de F\'{\i}sica, Universidad de
Extremadura, Badajoz, E-06071, Spain}
\author{Ren\'e D. Rohrmann}
\email{rohr@icate-conicet.gob.ar}
\homepage{http://icate-conicet.gob.ar/rohrmann}
\affiliation{Instituto de Ciencias Astron\'omicas, de la Tierra y del
Espacio (ICATE-CONICET), Avenida Espa\~na 1512 Sur, 5400 San Juan, Argentina}

\date{\today}

\begin{abstract}
The chemical potentials of multicomponent fluids are derived  in terms of the pair correlation functions for arbitrary number of components, interaction potentials, and dimensionality. The formally exact result is  particularized to hard-sphere mixtures with zero or positive nonadditivity. As a simple application, the chemical potentials of  three-dimensional additive hard-sphere mixtures are derived from the Percus--Yevick theory and the associated equation of state is obtained. This  Percus--Yevick chemical-route equation of state is shown to be more accurate than the virial equation of state. An interpolation  between the chemical-potential and compressibility routes  exhibits a better performance than the well-known Boubl\'ik--Mansoori--Carnahan--Starling--Leland equation of state.

\end{abstract}


\maketitle

\section{Introduction}
\label{sec1}
It is well known that in a fluid made of particles interacting via a pair-wise potential, the most relevant statistical-mechanical quantity in equilibrium is the pair correlation function,  or radial distribution function (RDF), $g(r)$ \cite{BH76,HM06}. It is {defined as the average number density at a distance $r$ from a certain reference particle, relative to the global density}. Apart from accounting for the structural correlation properties of the fluid, the RDF $g(r)$ allows one to obtain a number of thermodynamic quantities $\{\psi_i\}$ by means of elegant formulas. Since those thermodynamic quantities are connected by differential relations, one would expect to get the same macroscopic description, i.e., the same free energy $A$, regardless of the specific route $g(r)\to\psi_i\to A$ followed {\cite{note_13_03}}. However, this is not necessarily the case when an \emph{approximate} RDF is used as a starting point, what results in the well-known thermodynamic inconsistency problem \cite{BH76}.

To be more specific, let us assume an $s$-component system made of $N=\sum_{\alpha=1}^s N_\alpha$ particles ($N_\alpha$ being the number of particles of species $\alpha$) enclosed in a $d$-dimensional volume $V$.  The total number density is $\rho=N/V$ and the partial number densities are $\rho_\alpha=N_\alpha/V=x_\alpha\rho$, $x_\alpha=N_\alpha/N$ being the mole fractions. The pair interaction potential, {not necessarily isotropic,} between two particles of species $\alpha$ and $\gamma$ is $\phi_{\alpha\gamma}(\rr)=\phi_{\gamma\alpha}(-\rr)$. In terms of the pair correlation function of species $\alpha$ and $\gamma$, $g_{\alpha\gamma}(\rr)$, it is possible to express the  pressure $p$ as \cite{HM06}
\beq
Z\equiv \frac{\beta
p}{\rho}=1-\frac{\beta\rho}{2d}\sum_{\alpha,\gamma}x_\alpha x_\gamma\int \dd
\mathbf{r}\, g_{\alpha\gamma}(\mathbf{r})\mathbf{r}\cdot\nabla
\phi_{\alpha\gamma}(\mathbf{r}),
\label{1.1}
\eeq
where $Z$ is the compressibility factor and $\beta\equiv 1/k_BT$ ($k_B$ and $T$ being the Boltzmann constant
and the temperature, respectively).
Analogously, the internal energy per particle $u=U/N$ can be expressed as
\beq
\beta u=\frac{d}{2}+\frac{\beta\rho}{2}\sum_{\alpha,\gamma}x_\alpha x_\gamma\int \dd
\mathbf{r}\, g_{\alpha\gamma}(\mathbf{r})
\phi_{\alpha\gamma}(\mathbf{r}).
\label{1.2}
\eeq
Equations \eqref{1.1} and \eqref{1.2} are the virial (or pressure) and energy routes {to thermodynamics}, respectively. The free energy $A=Na$ (where $a$ is the free energy per particle) can be {(partially)} obtained {\cite{note_13_03}} either from Eq.\ \eqref{1.1} or from Eq.\ \eqref{1.2} by inversion of the thermodynamic relations
\beq
Z=\rho\left(\frac{\partial \beta a}{\partial\rho}\right)_{\beta,\{x_\alpha\}},
\label{1.3}
\eeq
\beq
u=\left(\frac{\partial \beta a}{\partial\beta}\right)_{\rho,\{x_\alpha\}}.
\label{1.4}
\eeq

Apart from the virial and energy routes, a popular route is the compressibility one. It reads
\begin{eqnarray}
\chi_{T}^{-1}&=&\left(\frac{\partial \rho Z}{\partial\rho}\right)_{\beta,\{x_\alpha\}}\nn
&=&
\sum_{\alpha,\gamma}\sqrt{x_\alpha x_\gamma}
\left(\mathsf{I}+\widehat{\mathsf{h}}\right)^{-1}_{\alpha\gamma},
\label{1.5}
\end{eqnarray}
where the element $\widehat{h}_{\alpha\gamma}$ of the matrix $\widehat{\mathsf{h}}$ is proportional to
the zero wavenumber limit of the Fourier transform of the total correlation function $h_{\alpha\gamma}(\mathbf{r})=g_{\alpha\gamma}(\mathbf{r})-1$, namely
\beq
\widehat{h}_{\alpha\gamma}=\rho\sqrt{x_\alpha x_\gamma}\int \dd\mathbf{r}\,h_{\alpha\gamma}\left(\mathbf{r}\right).
\label{1.6}
\eeq
Combination of Eqs.\ \eqref{1.3} and \eqref{1.5} allows one to get the free energy from the RDF via the compressibility route.

Equations \eqref{1.3} and \eqref{1.4} are related to the derivatives of the free energy per particle $a$ with respect to the total number density and to the temperature, respectively, but they ignore the composition-dependence of $a$. In fact, the partial derivatives of $a$ with respect to the partial number densities are related to the chemical potentials:
\beq
\mu_\nu=\left(\frac{\partial \rho a}{\partial\rho_\nu}\right)_{\beta,\{\rho_{\alpha\neq\nu}\}}.
\label{1.7}
\eeq
While in Eq.\ \eqref{1.3} the free energy per particle $a$ is seen as a function of the total number density $\rho$ and the $s-1$ independent mole fractions $\{x_1,x_2,\ldots, x_{s-1}\}$, in Eq.\ \eqref{1.7} $a$ is seen as a function of the $s$ partial densities $\{\rho_1,\rho_2,\ldots,\rho_s\}$. Adopting the former point of view, Eq.\ \eqref{1.7} can be rewritten as
\beq
\mu_\nu=\left(\frac{\partial\rho a}{\partial \rho}\right)_{\beta,\{x_\alpha\}}+\sum_{\alpha=1}^{s-1}\left(\frac{\partial a}{\partial x_\alpha}\right)_{\beta,\rho,\{x_{\gamma\neq\alpha}\}}\left(\delta_{\alpha\nu}-x_\alpha\right).
\label{1.8}
\eeq
This allows us to get the useful identity
\beq
\sum_\nu x_\nu \mu_\nu=\left(\frac{\partial\rho a}{\partial \rho}\right)_{\beta,\{x_\alpha\}}.
\label{1.9}
\eeq
Recognizing that $\left[\partial(\rho a)/{\partial \rho}\right]_{\beta,\{x_\alpha\}}=a+p/\rho$ is the Gibbs free energy per particle, Eq.\ \eqref{1.9} is not but the fundamental equation of thermodynamics \cite{C60}.

{Cross partial derivatives of the free energy should be independent of the order of derivation, which yields the following Maxwell relations involving the chemical potential:
\beq
\left(\frac{\partial \beta \mu_\nu}{\partial\beta}\right)_{\{\rho_{\alpha}\}}=\left(\frac{\partial \rho u}{\partial\rho_\nu}\right)_{\beta,\{\rho_{\alpha\neq\nu}\}},
\label{1.11}
\eeq
\beq
V\left(\frac{\partial \beta\mu_\nu}{\partial V}\right)_{\beta,\{N_{\alpha}\}}=-\left(\frac{\partial \rho Z}{\partial\rho_\nu}\right)_{\beta,\{\rho_{\alpha\neq\nu}\}},
\label{1.12}
\eeq
\beq
V^2\left(\frac{\partial^2 \beta\mu_\nu}{\partial V^2}\right)_{\beta,\{N_{\alpha}\}}=\left(\frac{\partial \rho\chi_T^{-1}}{\partial\rho_\nu}\right)_{\beta,\{\rho_{\alpha\neq\nu}\}},
\label{1.13}
\eeq
\beq
\left(\frac{\partial \mu_\nu}{\partial\rho_\alpha}\right)_{\beta,\{\rho_{\gamma\neq\alpha}\}}=\left(\frac{\partial \mu_\alpha}{\partial\rho_\nu}\right)_{\beta,\{\rho_{\gamma\neq\nu}\}}.
\label{1.10}
\eeq
In Eqs.\ \eqref{1.12} and \eqref{1.13} the replacement $\rho_\alpha\to N_\alpha/V$ must be done in order to express the chemical potential as a function of $\{N_\alpha\}$ and $V $.}

The interesting question is, how can one obtain the chemical potential $\mu_\nu$ (and hence the free energy) from the knowledge of the RDF? This question has been addressed in several textbooks \cite{H56,R80} and classical papers \cite{MR75,RFL59}, but the most general formula (valid for any dimensionality, interaction potential, and coupling protocol) seems not to have been derived yet. The first aim of this paper is the derivation of that chemical-potential route to thermodynamics [see Eq.\ \eqref{2.22} below]. The second goal is the application of the route to a mixture of (additive) hard spheres (HS) in the context of the Percus--Yevick (PY) approximation [see Eq.\ \eqref{5.3} below]. This extends to multicomponent systems, the work recently reported in Ref.\ \cite{S12b}.

{This paper is organized as follows. The formal derivation of the chemical-potential route is carried out in Sec.\ \ref{sec2}. As a simple test, it is checked in Sec.\ \ref{sec3} that the use of the exact pair correlation function to first order in density provides the exact second and third virial coefficients. Section \ref{sec4} is devoted to the specialization of the chemical-potential route to $d$-dimensional HS mixtures with zero or positive nonadditivity. In the three-dimensional case, the knowledge of the contact values of the radial distribution function for additive HS mixtures in the scaled-particle theory (SPT) and PY approximations is exploited to obtain the chemical potential. It is observed that, while the SPT yields a result thermodynamically consistent with the virial route, this is not the case of the PY result. Finally, the paper ends with some concluding remarks in Sec.\ \ref{sec6}.}

\section{The chemical-potential route}
\label{sec2}
As stated before, we consider an $s$-component mixture with $N_\alpha$ ($\alpha=1,\ldots,s$) particles of species $\alpha$ and $N=\sum_{\alpha=1}^s N_\alpha$ total number of particles in a volume $V$. We will employ the short-hand notations $\mathbf{N}\equiv\{N_1,\ldots,N_s\}$, $\mathbf{r}^N\equiv\{\mathbf{r}_1,\ldots, \mathbf{r}_N\}$, and $\dd\mathbf{r}^N\equiv \dd\mathbf{r}_1\cdots \dd\mathbf{r}_N$, $\mathbf{r}_i$ being the spatial coordinates of particle $i$. If we denote by $\Phi_\mathbf{N}(\mathbf{r}^N)$ the total potential energy, the (canonical-ensemble) configurational probability density is \cite{H56,HM06}
\beq
\rho_\mathbf{N}(\mathbf{r}^N)=\frac{V^{-N}}{Q_\NN(\beta,V)}e^{-\beta \Phi_\mathbf{N}(\mathbf{r}^N)},
\label{2.1}
\eeq
where
\beq
Q_\NN(\beta,V)=V^{-N}\int \dd\rN\,e^{-\beta \Phi_\mathbf{N}(\mathbf{r}^N)}
\label{2.2}
\eeq
is the configurational integral.
The average number of pairs of particles (per unit volume) of species $\alpha$ and $\gamma$  located at $\rr_\alpha$ and $\rr_\gamma$, respectively, is
\beqa
n_{\alpha\gamma}(\rr_\alpha,\rr_\gamma)&=&\int \dd\rN\, \rho_\NN(\rN)\sum_{i\neq j}\delta_{\epsilon_i,\alpha}
\delta_{\epsilon_j,\gamma}\nn
&&\times\delta(\rr_i-\rr_\alpha)\delta(\rr_j-\rr_\gamma),
\label{2.6}
\eeqa
where $\epsilon_i$ denotes the species  particle $i$ belongs to.
We define the pair correlation function as $g_{\alpha\gamma}(\rr_\alpha,\rr_\gamma)=n_{\alpha\gamma}(\rr_\alpha,\rr_\gamma)/\rho_\alpha\rho_\gamma$. Inserting Eq.\ \eqref{2.1} into Eq.\ \eqref{2.6}, one obtains
\beqa
g_{\alpha\gamma}(\rr_\alpha,\rr_\gamma)&=&\frac{V^{-(N-2)}}{Q_\NN(\beta,V)}\int \dd\rN\, e^{-\beta\Phi_\NN(\rN)}\nn
&&\times\delta(\rr_1-\rr_\alpha)\delta(\rr_2-\rr_\gamma),
\label{2.7}
\eeqa
where, without loss of generality, particles $i=1$ and $j=2$ are assumed to belong to species $\alpha$ and $\gamma$, respectively.

Contact with thermodynamics is made through the free energy $A_\NN$ of the system:
\beq
A_\NN(\beta,V)=-k_BT\ln \ZZ_\NN(\beta,V),
\label{2.3}
\eeq
where the partition function $\ZZ_\NN$ factorizes as
\beq
\ZZ_\NN(\beta,V)=\ZZ_\NN^\id(\beta,V) Q_\NN(\beta,V),
\label{2.4}
\eeq
\beq
\ZZ_\NN^\id(\beta,V)= \frac{V^N}{\prod_\alpha N_\alpha!\Lambda_\alpha^{dN_\alpha}}
\label{2.5}
\eeq
being the ideal-gas partition function. In Eq.\ \eqref{2.5},
$\Lambda_\alpha=h/\sqrt{2\pi m_\alpha k_BT}$ (where $h$ is the Planck constant and $m_\alpha$ is the mass of a particle of species $\alpha$) is the thermal de Broglie wavelength.

Let us now focus on a given species $\nu$. The associated chemical potential is
\beq
\mu_\nu=\frac{\partial A_\NN(\beta,V)}{\partial N_\nu}=\mu_\nu^\id+\mu_\nu^\ex,
\label{2.8}
\eeq
where
\beq
\beta \mu_\nu^\id=-\frac{\partial \ln \ZZ_\NN^\id(\beta,V)}{\partial N_\nu}=\ln\left(\rho_\nu\Lambda_\nu^d\right),
\eeq
\beqa
\beta \mu_\nu^\ex&=&-\frac{\partial \ln Q_\NN(\beta,V)}{\partial N_\nu}\nn
&=&\ln\frac{Q_\NN(\beta,V)}{Q_{\NN+1}(\beta,V)}.
\label{2.9}
\eeqa
In the second line of Eq.\ \eqref{2.9} we have taken into account that $N_\nu\gg 1$ and it is understood that the subscript $\NN+1$ means $\{N_1,\ldots,N_\nu+1,\ldots,N_s\}$. Obviously, $Q_{\NN+1}(\beta,V)$ is given by Eq.\ \eqref{2.2} with the replacements $N\to N+1$ and $\NN\to\NN+1$. Moreover, without loss of generality, we will assign the label $i=0$ to the extra particle of species $\nu$, so that $\dd\rr^{N+1}=\dd\rr_0d\rN$.

Now we assume that the potential energy is pair-wise additive, namely
\beq
\Phi_\NN(\rN)=\sum_{i=1}^{N-1}\sum_{j=i+1}^N\phi_{\epsilon_i\epsilon_j}(\rr_i,\rr_j),
\label{2.10}
\eeq
\beq
\Phi_{\NN+1}(\rr^{N+1})=\sum_{j=1}^{N}\phi_{\nu\epsilon_j}(\rr_0,\rr_j)+\Phi_\NN(\rN),
\label{2.11}
\eeq
where $\phi_{\alpha\gamma}(\rr_\alpha,\rr_\gamma)=\phi_{\alpha\gamma}(\rr_\gamma-\rr_\alpha)$ is the interaction potential of two particles of species $\alpha$ and $\gamma$.

In order to establish a relationship between the chemical potential and the pair correlation functions, it is convenient to introduce a \emph{coupling parameter} $\xi$ such that its value $0\leq\xi\leq 1$ controls the strength of the interaction of particle $i=0$ to the rest of particles. A similar charging process was employed by Onsager in a different context \cite{O33}. It is also closely related to the so-called Widom insertion method \cite{W63b}.

In our specific problem, the interaction potential between particles $i=0$ and $j\geq 1$ is $\phi_{\nu\epsilon_j}^\xx(\rr_0,\rr_j)$, with the boundary conditions
\beq
\phi_{\nu\epsilon_j}^\xx(\rr_0,\rr_j)=\begin{cases}
  0,&\xi=0,\\
  \phi_{\nu\epsilon_j}(\rr_0,\rr_j),&\xi=1.
\end{cases}
\label{2.12}
\eeq
The associated total potential energy and configuration integral are
\beq
\Phi_{\NN+1}^\xx(\rr^{N+1})=\sum_{j=1}^{N}\phi_{\nu\epsilon_j}^\xx(\rr_0,\rr_j)+\Phi_\NN(\rN),
\label{2.13}
\eeq
\beq
Q_{\NN+1}^\xx(\beta,V)=V^{-(N+1)}\int \dd\rr^{N+1}\,e^{-\beta \Phi_{\NN+1}^\xx(\rr^{N+1})}.
\label{2.14}
\eeq
Note that Eqs.\ \eqref{2.13} and \eqref{2.14} (with $0<\xi<1$) define a system of $N+1$ particles and $s+1$ species, one of the species being made by the single particle $i=0$. Only in the limits $\xi=0$ and $\xi=1$ does one recover a number  of species $s$. Analogously to Eq.\ \eqref{2.7}, the pair correlation function of particle $i=0$ and any particle of species $\alpha$ is
\beqa
g_{\nu\alpha}^\xx(\rr_\nu,\rr_\alpha)&=&\frac{V^{-(N-1)}}{Q_{\NN+1}^\xx(\beta,V)}\int \dd\rr^{N+1}\, e^{-\beta\Phi_{\NN+1}^\xx(\rr^{N+1})}\nn
&&\times\delta(\rr_0-\rr_\nu)\delta(\rr_1-\rr_\alpha).
\label{2.17}
\eeqa
Obviously,
\beq
\Phi_{\NN+1}^\xx(\rr^{N+1})=\begin{cases}
  \Phi_{\NN}(\rr^{N}),&\xi=0\\
  \Phi_{\NN+1}(\rr^{N+1}),&\xi=1,
\end{cases}
\label{2.15}
\eeq
\beq
Q_{\NN+1}^\xx(\beta,V)=\begin{cases}
  Q_{\NN}(\beta,V),&\xi=0\\
  Q_{\NN+1}(\beta,V),&\xi=1,
\end{cases}
\label{2.16}
\eeq
\beq
g_{\nu\alpha}^\xx(\rr_\nu,\rr_\alpha)=
\begin{cases}
  1,&\xi=0\\
 g_{\nu\alpha}(\rr_\nu,\rr_\alpha),&\xi=1.
\end{cases}
\label{2.18}
\eeq

Assuming that the potentials $\phi_{\nu\alpha}^\xx$ are differentiable with respect to $\xi$, the second line of Eq.\ \eqref{2.9} can be rewritten as
\beq
\beta \mu_\nu^\ex=-\int_0^1 \dd\xi\, \frac{\partial}{\partial \xi}\ln Q_{\NN+1}^\xx(\beta,V).
\label{2.19}
\eeq
Now, from Eqs.\ \eqref{2.13} and \eqref{2.14} we obtain
\beqa
\partial_\xi Q_{\NN+1}^\xx&=&-\beta V^{-(N+1)}\int \dd\rr^{N+1}\, e^{-\beta \Phi_{\NN+1}^\xx(\rr^{N+1})}\nn
&&\times\sum_{j=1}^N \partial_\xi
\phi_{\nu\epsilon_j}^\xx(\rr_0,\rr_j)\nn
&=&-\beta V^{-N}\sum_\alpha \rho_\alpha
\int \dd\rr^{N+1}\, e^{-\beta \Phi_{\NN+1}^\xx(\rr^{N+1})}\nn
&&\times \partial_\xi
\phi_{\nu\alpha}^\xx(\rr_0,\rr_1),
\label{2.20}
\eeqa
where again particle $j=1$ is assumed (without loss of generality) to belong to species $\alpha$.
Making use of Eq.\ \eqref{2.17},
\beqa
\partial_\xi \ln Q_{\NN+1}^\xx&=&-\frac{\beta}{V}\sum_\alpha \rho_\alpha\int \dd\rr_\nu\int \dd\rr_\alpha\,g_{\nu\alpha}^\xx(\rr_\nu,\rr_\alpha)\nn
&&\times\partial_\xi
\phi_{\nu\alpha}^\xx(\rr_\nu,\rr_\alpha).
\label{2.21}
\eeqa
Combination of Eqs.\ \eqref{2.19} and \eqref{2.21} yields the desired result
\beq
\mu_\nu^\ex=\sum_\alpha \rho_\alpha\int_0^1 \dd\xi \int \dd\rr\, g_{\nu\alpha}^\xx(\rr)\frac{\partial \phi_{\nu\alpha}^\xx(\rr)}{\partial\xi},
\label{2.22}
\eeq
where we have taken into account the translational invariance property $\phi_{\nu\alpha}^\xx(\rr_\nu,\rr_\alpha)=\phi_{\nu\alpha}^\xx(\rr_\alpha-\rr_\nu)$.
In terms of the cavity function $y_{\nu\alpha}^\xx(\rr)=g_{\nu\alpha}^\xx(\rr)e^{\beta \phi_{\nu\alpha}^\xx(\rr)}$, Eq.\ \eqref{2.22} becomes
\beq
\beta\mu_\nu^\ex=-\sum_\alpha \rho_\alpha\int_0^1 \dd\xi \int \dd\rr\, y_{\nu\alpha}^\xx(\rr)\frac{\partial e^{-\beta \phi_{\nu\alpha}^\xx(\rr)}}{\partial\xi}.
\label{2.23}
\eeq

Equation \eqref{2.22} or, equivalently, Eq.\ \eqref{2.23} constitutes the chemical-potential route to thermodynamics. It expresses the chemical potential of species $\nu$ in an $s$-component mixture (the ``solvent'') in terms of the pair correlations of an $(s+1)$-component mixture where the extra component consists of a single particle (the ``solute'') whose interaction with the rest of the particles is controlled by a coupling parameter $\xi$, which switches from $\xi=0$ (the solute {does not feel the presence of the solvent particles}) to $\xi=1$ (the solute is indistinguishable from a {solvent} particle of species $\nu$). The result is independent of the number  of species $s$, the dimensionality  of the system $d$, the pair interaction functions $\phi_{\alpha\gamma}(\rr)$ (which can be isotropic or not), and the protocol $\phi_{\nu\alpha}^\xx(\rr)$ followed to go from $\phi_{\nu\alpha}^{(0)}(\rr)=0$ to $\phi_{\nu\alpha}^{(1)}(\rr)=\phi_{\nu\alpha}(\rr)$. If the exact $g_{\nu\alpha}^\xx(\rr)$ is employed, one gets the exact chemical potential $\mu_\nu$ (and, hence, the exact free energy $A$) regardless of the protocol. This is illustrated in Sec.\ \ref{sec3} at the level of the third virial coefficient. On the other hand, if an approximate function $g_{\nu\alpha}^\xx(\rr)$ is used instead, not only the resulting approximate free energy will be different from the one derived from any of the other routes (e.g., virial, energy, and compressibility), but it will, in general, depend on the choice of the protocol. Thus, the chemical-potential route adds an extra source of thermodynamic inconsistency.

\section{Low-density regime}
\label{sec3}
The aim of this section is to exploit Eq.\ \eqref{2.23} when the pair correlation function is truncated to first order in density.
First, let us rewrite Eq.\ \eqref{2.23} as
\beqa
\beta\mu_\nu^\ex&=&-V^{-1}\int_0^1 \dd\xi\,\sum_\alpha \rho_\alpha \int \dd\rr_\nu\int \dd\rr_\alpha\, y_{\nu\alpha}^\xx(\rr_\nu,\rr_\alpha)\nn
&&\times\partial_\xi f_{\nu\alpha}^\xx(\rr_\nu,\rr_\alpha),
\label{3.1}
\eeqa
where $f_{\nu\alpha}^\xx(\rr_\nu,\rr_\alpha)\equiv e^{-\beta \phi_{\nu\alpha}^\xx(\rr_\nu,\rr_\alpha)}-1$ is the Mayer function \cite{HM06,H56}.

To first order in density,
\beqa
y_{\nu\alpha}^\xx(\rr_\nu,\rr_\alpha)&=&1+\sum_{\gamma}\rho_\gamma\int \dd\rr_\gamma\,f_{\nu\gamma}^\xx(\rr_\nu,\rr_\gamma)f_{\gamma\alpha}(\rr_\gamma,\rr_\alpha)\nn
&&+\mathcal{O}(\rho^2),
\label{3.2}
\eeqa
where $f_{\gamma\alpha}(\rr_\gamma,\rr_\alpha)\equiv e^{-\beta \phi_{\gamma\alpha}(\rr_\gamma,\rr_\alpha)}-1$. Insertion into Eq.\ \eqref{3.1} yields
\beqa
\beta\mu_\nu^\ex&=&-V^{-1}\sum_\alpha \rho_\alpha \int \dd\rr_\nu\int \dd\rr_\alpha\, f_{\nu\alpha}(\rr_\nu,\rr_\alpha)\nn
&&-V^{-1}\int_0^1 \dd\xi\,I_\nu^\xx+\mathcal{O}(\rho^3),
\label{3.3}
\eeqa
where we have called
\beqa
I_\nu^\xx&\equiv &\sum_{\alpha,\gamma} \rho_\alpha\rho_\gamma \int \dd\rr_\nu\int \dd\rr_\alpha\int \dd\rr_\gamma\,
f_{\nu\gamma}^\xx(\rr_\nu,\rr_\gamma)f_{\gamma\alpha}(\rr_\gamma,\rr_\alpha)\nn
&&\times\partial_\xi f_{\nu\alpha}^\xx(\rr_\nu,\rr_\alpha).
\label{3.4}
\eeqa
Exchanging $\alpha\leftrightarrow\gamma$ and $\rr_\alpha\leftrightarrow\rr_\gamma$ in the summation and the integral,
$I_\nu^\xx$ can be rewritten as
\beqa
I_\nu^\xx&=&\frac{1}{2}\sum_{\alpha,\gamma} \rho_\alpha\rho_\gamma \int \dd\rr_\nu\int \dd\rr_\alpha\int \dd\rr_\gamma\,
f_{\gamma\alpha}(\rr_\gamma,\rr_\alpha)\nn
&&\times \partial_\xi\left[f_{\nu\alpha}^\xx(\rr_\nu,\rr_\alpha)f_{\nu\gamma}^\xx(\rr_\nu,\rr_\gamma)\right].
\label{3.5}
\eeqa
Therefore, Eq.\ \eqref{3.3} becomes
\beq
\beta\mu_\nu^\ex=-\sum_\alpha \rho_\alpha \mathcal{F}_{\nu\alpha}-\frac{1}{2}\sum_{\alpha,\gamma} \rho_\alpha\rho_\gamma \mathcal{G}_{\nu\alpha\gamma}+\mathcal{O}(\rho^3),
\label{3.6}
\eeq
where
\beq
\mathcal{F}_{\nu\alpha}\equiv \int \dd\rr\, f_{\nu\alpha}(\rr),
\label{3.7}
\eeq
\beqa
\mathcal{G}_{\nu\alpha\gamma}&\equiv& \int \dd\rr\int \dd\rr'\,
f_{\gamma\alpha}(\rr-\rr') f_{\nu\alpha}(\rr)f_{\nu\gamma}(\rr'),
\label{3.8}
\eeqa
and the translational invariance property has been applied again.
Note that both $\mathcal{F}_{\nu\alpha}$ and $\mathcal{G}_{\nu\alpha\gamma}$ are invariant under any permutation of indices. This guarantees that Eq.\ \eqref{1.10} is verified.

Equations \eqref{3.6}--\eqref{3.8} show that, as expected, the specific protocol does not play any role in the final result.
{}From Eq.\ \eqref{1.7} we finally get
\beq
\beta\rho a^\ex=-\frac{1}{2}
\sum_{\alpha,\gamma} \rho_\alpha\rho_\gamma \mathcal{F}_{\alpha\gamma}-\frac{1}{6}\sum_{\alpha,\gamma,\nu} \rho_\alpha\rho_\gamma \rho_\nu \mathcal{G}_{\alpha\gamma\nu}+\mathcal{O}(\rho^4).
\label{3.9}
\eeq
Making use of Eq.\ \eqref{1.3} it is straightforward to get
\beq
Z=1+B_2\rho+B_3\rho^2+\mathcal{O}(\rho^3),
\label{3.10}
\eeq
where the second and third virial coefficients are
\beq
B_2=-\frac{1}{2}\sum_{\alpha,\gamma} x_\alpha x_\gamma \mathcal{F}_{\alpha\gamma},
\label{3.11}
\eeq
\beq
B_3=-\frac{1}{3}\sum_{\alpha,\gamma,\nu} x_\alpha x_\gamma x_\nu \mathcal{G}_{\alpha\gamma\nu}.
\label{3.12}
\eeq
{As expected, these are the exact expressions \cite{HM06}.}

\section{Hard-sphere mixtures\label{sec4}}

Let us now particularize the chemical-potential route \eqref{2.23} to  HS mixtures. In that case,
\beq
e^{-\beta\phi_{\alpha\gamma}(\rr)}=\Theta(r-\sigma_{\alpha\gamma}),
\label{4.1}
\eeq
where $r=|\mathbf{r}|$, $\Theta(x)$ is Heaviside's step function, and $\sigma_{\alpha\gamma}$ is the range of the infinitely repulsive interaction between particles of species $\alpha$ and $\gamma$.

We now need to introduce the extra particle $i=0$ (the solute) coupled to the remaining $N$ particles through a coupling parameter $0\leq\xi\leq 1$ via the set of interaction potentials $\phi_{\nu\alpha}^\xx(\rr)$. According to Eq.\ \eqref{2.12}, $\exp[-\beta \phi_{\nu\alpha}^{(0)}(\rr)]=1$ and $\exp[-\beta \phi_{\nu\alpha}^{(1)}(\rr)]=\Theta(r-\sigma_{\nu\alpha})$, but otherwise a certain freedom to fix the protocol $\phi_{\nu\alpha}^\xx(\rr)$ exists. The most natural choice is a HS form, i.e.,
\beq
e^{-\beta\phi_{\nu\alpha}^\xx(\rr)}=\Theta(r-\sigma_{\nu\alpha}^\xx),
\label{4.2}
\eeq
with $\sigma_{\nu\alpha}^{(0)}=0$ and $\sigma_{\nu\alpha}^{(1)}=\sigma_{\nu\alpha}$. Therefore,
\beq
\frac{\partial e^{-\beta\phi_{\nu\alpha}^\xx(\rr)}}{\partial\xi}=-\delta(r-\sigma_{\nu\alpha}^\xx)\frac{\partial \sigma_{\nu\alpha}^\xx}{\partial\xi},
\eeq
so that Eq.\ \eqref{2.23} becomes
\beq
\beta\mu_\nu^\ex=d 2^dv_d\sum_\alpha \rho_\alpha\int_{0}^{\sigma_{\nu\alpha}} \dd\sigma_{0\alpha} \,\sigma_{0\alpha}^{d-1}y_{0\alpha}(\sigma_{0\alpha}),
\label{4.10}
\eeq
where we have taken into account that the integral of $\dd\rr$ over all orientations is $d 2^d v_d r^{d-1}\dd r$,  $v_{d}=(\pi /4)^{d/2}/\Gamma (1+d/2)$ being the volume of a
$d$-dimensional sphere of unit diameter. Note that in Eq.\ \eqref{4.10} the integration variable has changed from $\xi$ to $\sigma_{\nu\alpha}^\xx$ and we have simplified the notation as $\sigma_{\nu\alpha}^\xx\to \sigma_{0\alpha}$ and $y_{\nu\alpha}^\xx\to y_{0\alpha}$. This change avoids the need to specify the $\xi$-dependence of $\sigma_{\nu\alpha}^\xx$.

{Equation \eqref{4.10} applies to any choice of the set $\{\sigma_{\alpha\gamma}\}$.
Henceforth we will assume the  condition $\sigma_{\alpha\gamma}\geq\frac{1}{2}(\sigma_\alpha+\sigma_\gamma)$, where $\sigma_\alpha$ and $\sigma_\gamma$ are the diameters of particles of species $\alpha$ and $\gamma$, respectively. This implies that the HS mixture is either additive or has a positive nonadditivity.
In such a case, Eq.\ \eqref{4.10} can be further simplified. To that end, it} is convenient to decompose $\beta\mu_\nu^\ex$ into two pieces,
\beq
\beta\mu_\nu^\ex=\beta\mu_\nu^{\ex,\text{I}}+\beta\mu_\nu^{\ex,\text{II}},
\eeq
where
\beq
\beta\mu_\nu^{\ex,\text{I}}=d 2^dv_d\sum_\alpha \rho_\alpha\int_{0}^{\frac{1}{2}\sigma_{\alpha}} \dd\sigma_{0\alpha} \,\sigma_{0\alpha}^{d-1}y_{0\alpha}(\sigma_{0\alpha}),
\label{4.10.I}
\eeq
\beq
\beta\mu_\nu^{\ex,\text{II}}=d 2^dv_d\sum_\alpha \rho_\alpha\int_{\frac{1}{2}\sigma_{\alpha}}^{\sigma_{\nu\alpha}} \dd\sigma_{0\alpha} \,\sigma_{0\alpha}^{d-1}y_{0\alpha}(\sigma_{0\alpha}).
\label{4.10.II}
\eeq
In the first contribution, the condition $\sigma_{0\alpha}\leq \frac{1}{2}\sigma_{\alpha}$  implies that the solute  can lie ``inside'' a particle of species $\alpha$. This makes the contribution $\beta\mu_\nu^{\ex,\text{I}}$ very easy to evaluate. First, by reversing the steps leading from Eq.\ \eqref{2.9} to Eq.\ \eqref{4.10}, one can write
\beq
\beta\mu_\nu^{\ex,\text{I}}=
\ln\frac{Q_\NN(\beta,V)}{Q_{\NN+1}^{(\xi_0)}(\beta,V)},
\eeq
where $\xi_0$ denotes the \emph{common} value of the coupling parameter $\xi$ at which $\sigma_{\nu\alpha}^\xx=\sigma_{0\alpha}$ takes the value $\frac{1}{2}\sigma_\alpha$ for all $\alpha$.
According to Eqs.\ \eqref{2.13}, \eqref{2.14}, and \eqref{4.2},
\beqa
Q_{\NN+1}^{(\xi_0)}(\beta,V)&=&V^{-(N+1)}\int \dd\rr^{N}\,e^{-\beta \Phi_{\NN}(\rr^{N})}\nn
&&\times\int \dd\rr_0\,\prod_{j=1}^N
\Theta\left(|\rr_0-\rr_j|-\frac{1}{2}\sigma_{\epsilon_j}\right).
\label{4.5}
\eeqa
Next, thanks to the condition $\sigma_{\alpha\gamma}\geq \frac{1}{2}(\sigma_\alpha+\sigma_\gamma)$,  we note that in Eq.\ \eqref{4.5} the spatial regions excluded to particle $i=0$ by the remaining particles $j=1,\ldots,N$ do not overlap, so that
\beq
\int \dd\rr_0\,\prod_{j=1}^N
\Theta\left(|\rr_0-\rr_j|-\frac{1}{2}\sigma_{\epsilon_j}\right)=V(1-\eta),
\label{4.7}
\eeq
where  $\eta=v_d \sum_{\alpha}\rho_\alpha\sigma_\alpha^d$ is the total packing fraction of the solvent system. Thus,  Eq.\ \eqref{4.5} reduces to
$
Q_{\NN+1}^{(\xi_0)}(\beta,V)=Q_{\NN}(\beta,V)(1-\eta)$ and therefore $\beta\mu_\nu^{\ex,\text{I}}=-\ln(1-\eta)$.

Taking all of this into account, we finally get
\beq
\beta\mu_\nu^\ex=-\ln(1-\eta)+d 2^dv_d\sum_\alpha \rho_\alpha\int_{\frac{1}{2}\sigma_{\alpha}}^{\sigma_{\nu\alpha}} \dd\sigma_{0\alpha} \,\sigma_{0\alpha}^{d-1} y_{0\alpha}(\sigma_{0\alpha}).
\label{4.10b}
\eeq

\section{Application to hard-sphere mixtures in the PY and SPT approximations}
\label{sec5}
In this section, we apply Eq.\ \eqref{4.10b} to the derivation of the chemical potential of a three-dimensional \emph{additive} HS fluid mixture, i.e., $\sigma_{\alpha\gamma}= \frac{1}{2}(\sigma_\alpha+\sigma_\gamma)$, as resulting from  the PY and SPT approximations. Given an $s$-component mixture, the corresponding contact values are
\beqa
y_{\alpha\gamma}(\sigma_{\alpha\gamma})&=&\frac{1}{1-\eta
}+\frac{3}{2}\frac{\eta }{(1-\eta )^{2}}\frac{\sigma_\alpha\sigma_\gamma
{M}_2}{\sigma_{\alpha\gamma}{M}_3}\nn
&&+q\frac{\eta^{2}}{(1-\eta)^{3}}\left(\frac{\sigma_\alpha\sigma_\gamma
{M}_2}{\sigma_{\alpha\gamma}{M}_3}\right)^{2},
\label{5.1}
\eeqa
where
\beq
M_n\equiv \sum_{\alpha=1}^s x_\alpha\sigma_\alpha^n,
\eeq
and the parameter $q$ takes the values $q=0$ and $q=\frac{3}{4}$ for the PY \cite{L64} and SPT \cite{RFL59,HFL61,LHP65,MR75,R88,HC04} theories, respectively. The more accurate Boubl\'ik--Grundke--Henderson--Lee--Levesque (BGHLL) \cite{B70,GH72,LL73} expression corresponds to the intermediate value $q=\frac{1}{2}$.

Now, in order to apply Eq.\ \eqref{4.10b}, we assume that a solute particle $i=0$ is introduced in such a way that it interacts with a particle of species $\alpha$ through a HS potential of range $\sigma_{0\alpha}$,  the contact value of the corresponding cavity function being $y_{0\alpha}(\sigma_{0\alpha})$. In principle, we are free to choose $\sigma_{0\alpha}$ within the interval $\frac{1}{2}\sigma_\alpha\leq \sigma_{0\alpha}\leq \sigma_{\nu\alpha}$. On the other hand, if we want to make use of the approximation \eqref{5.1}, we need to make the solute particle interact additively with the rest of the particles in the system. This is achieved by assuming that the solute particle is a sphere of diameter $\sigma_0$ within the range $0\leq\sigma_0\leq \sigma_\nu$ and $\sigma_{0\alpha}=\frac{1}{2}(\sigma_0+\sigma_\alpha)$. In that case, Eq.\ \eqref{4.10b} becomes
\beq
\beta\mu_\nu^\ex=-\ln(1-\eta)+\frac{12\eta}{M_3}\sum_\alpha x_\alpha\int_{0}^{\sigma_\nu} \dd\sigma_{0} \,\sigma_{0\alpha}^2 y_{0\alpha}(\sigma_{0\alpha}),
\label{5.2}
\eeq
where $y_{0\alpha}(\sigma_{0\alpha})$ is simply given by Eq.\ \eqref{5.1} with $\sigma_\gamma\to\sigma_0$. After simple algebra, Eq.\ \eqref{5.2} yields
\beqa
\beta\mu_\nu^\ex&=&-\ln(1-\eta)+\frac{3\eta}{1-\eta}\frac{M_2}{M_3}\left\{\sigma_\nu+\left[\frac{M_1}{M_2}+\frac{3\eta}{2(1-\eta)}
\right.\right.\nn
&&\left.\times
\frac{M_2}{M_3}\right]\sigma_\nu^2+\left[\frac{1}{3M_2}+\frac{\eta}{1-\eta}\frac{M_1}{M_3}+\frac{4q\eta^2}{3(1-\eta)^2}
\right.\nn
&&\left.\left.\times\frac{M_2^2}{M_3^2}
\right]\sigma_\nu^3\right\}.
\label{5.3}
\eeqa

Equation \eqref{5.3} shows that  $\beta\mu_\nu^\ex$ is an explicit function of $\sigma_\nu$ and depends on the diameters of all the species and on the partial densities through $\eta$, $M_1$, $M_2$, and $M_3$, or equivalently, through $\zeta_n\equiv \sum_\alpha \rho_\alpha\sigma_\alpha^n=\rho M_n$ with $n=0$--$3$. In terms of the latter quantities, Eq.\ \eqref{5.3} becomes
\beqa
\beta\mu_\nu^\ex&=&-\ln(1-\eta)+\frac{\pi}{2}\frac{\zeta_2}{1-\eta}\sigma_\nu+\left[
\frac{\pi}{2}\frac{\zeta_1}{1-\eta}\right.\nn
&&\left.+\frac{\pi^2}{8}\frac{\zeta_2^2}{(1-\eta)^2}\right]\sigma_\nu^2+\left[\frac{\pi}{6}\frac{\zeta_0}{1-\eta}
+\frac{\pi^2}{12}\frac{\zeta_1\zeta_2}{(1-\eta)^2}\right.\nn
&&\left.+q\frac{\pi^3}{54}\frac{\zeta_2^3}{(1-\eta)^3}
\right]\sigma_\nu^3,
\label{5.3b}
\eeqa
where one must take into account that $\eta=\frac{\pi}{6}\zeta_3$.
The next step would be to derive the excess free energy per particle, $a^\ex$, from application of Eq.\ \eqref{1.7}. Nevertheless, it turns out that the free energy associated with Eqs.\ \eqref{5.3} or \eqref{5.3b} is not well defined in the multicomponent case unless $q=\frac{3}{4}$. This is because  from Eq.\ \eqref{5.3b} we get
\beqa
\left(\frac{\partial \beta\mu_\nu}{\partial\rho_\alpha}\right)_{\{\rho_{\gamma\neq\alpha}\}}&-&\left(\frac{\partial \beta\mu_\alpha}{\partial\rho_\nu}\right)_{\{\rho_{\gamma\neq\nu}\}}=\frac{\pi^3}{24}\frac{\zeta_2^3}{(1-\eta)^3}\sigma_\alpha^2\sigma_\nu^2
\nn
&&\times\left(\sigma_\alpha-\sigma_\nu\right)
\left(1-\frac{4q}{3}\right),
\label{5.4F}
\eeqa
whereas Eq.\ \eqref{1.7} implies the (Maxwell) symmetry relation \eqref{1.10}.
Therefore, except in the one-component case ($\sigma_\alpha=\sigma_\nu$), ${\partial \mu_\nu}/{\partial\rho_\alpha}\neq{\partial \mu_\alpha}/{\partial\rho_\nu}$ unless $q=\frac{3}{4}$.
Therefore, the PY prescription \eqref{5.1} (with $q=0$), when inserted into the chemical-potential route \eqref{4.10b}, gives an expression for the chemical potential of the mixture that is not strictly consistent with a well-defined free energy. On the other hand,  this difficulty can be circumvented by using Eq.\ \eqref{1.9} instead of Eq.\ \eqref{1.7} to obtain $\left[\partial(\beta\rho a^\ex)/{\partial \rho}\right]_{\{x_\alpha\}}$. Integration over density (with the integration constant fixed by the condition $\lim_{\rho\to 0}a^\ex=0$) then yields
\beqa
\beta a^\ex&=&-\ln(1-\eta)+\frac{3\eta}{1-\eta}\frac{M_1M_2}{M_3}
+\frac{3\eta^2}{2(1-\eta)^2}\frac{M_2^3}{M_3^2}\nn
&&+
\left(1-\frac{4q}{3}\right)\frac{3M_2^3}{2M_3^2}\left[\frac{6-9\eta+2\eta^2}{(1-\eta)^2}
+6\frac{\ln(1-\eta)}{\eta}\right].\nn
\label{5.6}
\eeqa
{If now the chemical potential is obtained from Eq.\ \eqref{5.6} via Eq.\ \eqref{1.7}, the result differs from Eq.\ \eqref{5.3} unless $q=\frac{3}{4}$, namely
\beqa
\left(\frac{\partial \beta\rho a^\ex}{\partial\rho_\nu}\right)_{\beta,\{\rho_{\alpha\neq\nu}\}}&=&\beta\mu_\nu^\ex+
\left(1-\frac{4q}{3}\right)\frac{9M_2^2}{2M_3^2}\nn
&&\times \left[\frac{6-9\eta+2\eta^2}{(1-\eta)^2}
+6\frac{\ln(1-\eta)}{\eta}\right]\nn
&&\times\left(\sigma_\nu^2-\frac{M_2}{M_3}\sigma_\nu^3\right).
\label{new}
\eeqa
Of course, although $\left({\partial \rho a^\ex}/{\partial\rho_\nu}\right)_{\beta,\{\rho_{\alpha\neq\nu}\}}\neq\mu_\nu^\ex$, one has $\sum_\nu x_\nu \left({\partial\rho a^\ex}/{\partial\rho_\nu}\right)_{\beta,\{\rho_{\alpha\neq\nu}\}}=\sum_\nu x_\nu\mu_\nu^\ex$, so that Eq.\ \eqref{5.6} is recovered  from both alternative expressions for the chemical potential via Eq.\ \eqref{1.9}. Obviously, the right-hand side of Eq.\ \eqref{new} satisfies the symmetry condition $\partial^2 \rho a^\ex/\partial\rho_\alpha\partial\rho_\nu=\partial^2 \rho a^\ex/\partial\rho_\nu\partial\rho_\alpha$.}

The  equation of state corresponding to the free energy \eqref{5.6} is obtained from Eq.\ \eqref{1.3} as
\beqa
Z_\mu&=&\frac{1}{1-\eta}+\frac{3\eta}{(1-\eta)^2}\frac{M_1M_2}{M_3}
+\frac{3\eta^2}{(1-\eta)^3}\frac{M_2^3}{M_3^2}\nn
&&-
\left(1-\frac{4q}{3}\right)\frac{3M_2^3}{2M_3^2}\left[\frac{6-15\eta+11\eta^2}{(1-\eta)^3}
+6\frac{\ln(1-\eta)}{\eta}\right],\nn
\label{5.7}
\eeqa
where we have taken into account that the ideal-gas contribution is $\lim_{\rho\to 0}Z=1$. In Eq.\ \eqref{5.7} the subscript $\mu$ has been introduced to emphasize that this equation of state has been derived from the chemical-potential route.

Let us contrast Eq.\ \eqref{5.7} with the equation of state obtained from the virial route, Eq.\ \eqref{1.1}. In the case of a (three-dimensional) HS fluid, Eq.\ \eqref{1.1} becomes
\beq
Z=1+\frac{4\eta}{M_3}\sum_{\alpha,\gamma}x_\alpha x_\gamma \sigma_{\alpha\gamma}^3 y_{\alpha\gamma}(\sigma_{\alpha\gamma}).
\label{5.8}
\eeq
Insertion of Eq.\ \eqref{5.1} gives
\beqa
Z_v&=&\frac{1}{1-\eta}+\frac{3\eta}{(1-\eta)^2}\frac{M_1M_2}{M_3}
+\frac{3\eta^2}{(1-\eta)^3}\frac{M_2^3}{M_3^2}\nn
&&\times\left[1-\left(1-\frac{4q}{3}\right)\eta\right],
\label{5.9}
\eeqa
where the subscript $v$ refers to the use of the virial route.
Comparison between Eqs.\ \eqref{5.7} and \eqref{5.9} shows that the chemical-potential and virial routes coincide in the SPT ($q=\frac{3}{4}$) but not in the PY ($q=0$) or BGHLL ($q=\frac{1}{2}$) approaches. Interestingly enough, the SPT equation of state coincides with that derived from the PY solution \cite{L64} via the compressibility route \eqref{1.5}.

To summarize, the three PY equations of state as obtained from the chemical-potential, virial, and compressibility routes are
\beq
Z_{\py \mu}=Z_\mu(q=0),
\label{5.10}
\eeq
\beq
Z_{\py v}=Z_v(q=0),
\label{5.11}
\eeq
\beq
Z_{\py c}=Z_v\left(q=\frac{3}{4}\right),
\label{5.12}
\eeq
respectively.

The celebrated Boubl\'ik--Mansoori--Carnahan--Starling--Leland (BMCSL) equation \cite{B70,MCSL71} is obtained as an interpolation between the $\py v$ and the $\py c$ equations of state with respective weights $\frac{1}{3}$ and $\frac{2}{3}$, i.e.,
\beqa
Z_{\text{BMCSL}}&=&\frac{1}{3}Z_{\py v}+\frac{2}{3}Z_{\py c}\nn
&=&Z_v\left(q=\frac{1}{2}\right).
\label{5.13}
\eeqa
Having a third PY equation of state, Eq.\ \eqref{5.10}, it seems natural to construct an alternative interpolation formula as
\beq
Z_{\py \mu c}=\alpha Z_{\py \mu}+(1-\alpha)Z_{\py c}.
\label{5.14}
\eeq
In the one-component case \cite{S12b}, values of the interpolation parameter $\alpha\approx 0.4$ were seen to provide a better equation of state than the standard Carnahan--Starling equation \cite{CS69}. In particular, the values $\alpha=\frac{2}{5}$ and $\alpha=\frac{7}{18}$ were explicitly considered.

\begin{figure}[htb]
  \includegraphics[width=.9\columnwidth]{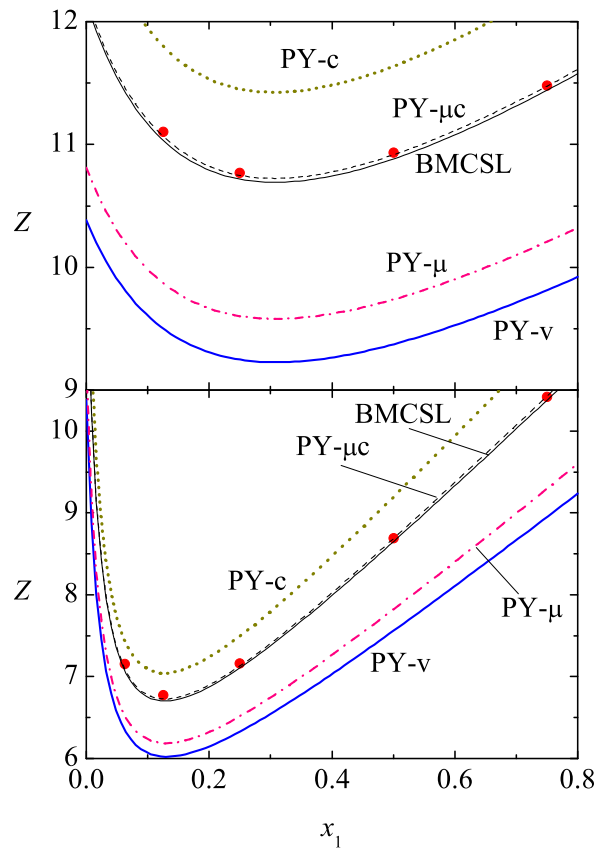}
\caption{(Color online). Plot of the compressibility factor $Z$ as a function of the mole fraction $x_1$ for a binary mixture with a packing fraction $\eta=0.49$ and a size ratio $\sigma_2/\sigma_1=0.6$ (top panel) or $\sigma_2/\sigma_1=0.3$ (bottom panel). The symbols are computer simulation values \protect\cite{BMLS96}, while the lines stand for (from top to bottom) Eqs.\ \protect\eqref{5.12}, \protect\eqref{5.14} with $\alpha=0.37$, \protect\eqref{5.13}, \protect\eqref{5.10}, and \protect\eqref{5.11}, respectively.\label{fig}}
\end{figure}

In order to assess the performance of Eqs.\ \eqref{5.10}--\eqref{5.14} against computer simulations for binary mixtures, we have chosen the highest packing fraction ($\eta=0.49$) and the two largest size disparities ($\sigma_2/\sigma_1=0.6$ and $\sigma_2/\sigma_1=0.3$) considered in Ref.\ \cite{BMLS96}. The simulation and theoretical results are displayed in Fig.\ \ref{fig}.
It is observed that, as expected, $Z_{\py v}$  underestimates the simulation values, while $Z_{\py c}$ overestimates them. The chemical-potential route $Z_{\py \mu}$ lies below the simulation data, but it exhibits a better  behavior than the virial route $Z_{\py v}$. The weighted average between $Z_{\py v}$ and $Z_{\py c}$ made in the construction of the BMCSL equation of state \eqref{5.13} does a very good job. A slightly better agreement is obtained  from the weighted average between $Z_{\py \mu}$ and $Z_{\py c}$ [cf.\ Eq.\ \eqref{5.14}] with $\alpha=\frac{2}{5}$ or $\alpha=\frac{7}{18}$ (not shown) but a value $\alpha=0.37$ provides especially accurate results.

\section{Conclusions}
\label{sec6}
{
In this paper we have revisited the problem on the derivation of the chemical potential of a fluid system at equilibrium from the knowledge of the pair correlation functions. The result, Eqs.\ \eqref{2.22} or \eqref{2.23}, applies to  a mixture with any number of components and generic interaction potentials $\phi_{\alpha\gamma}(\rr)$ between particles of species $\alpha$ and $\gamma$. The obtention of the chemical potential of species $\nu$ relies on the formal introduction of an extra particle (the solute) coupled to the rest of the particles of the system (the solvent) via a  series of interaction potentials $\phi_{\nu\alpha}^\xx(\rr)$, which depend on a certain charging parameter $\xi$ in such a way that at $\xi=0$ the solute ignores the presence of the solvent particle, while at $\xi=1$ the solute becomes indistinguishable from a solvent particle of species $\nu$ [see Eq.\ \eqref{2.12}]. The choice of the protocol leading from $\phi_{\nu\alpha}^\xx(\rr)=0$ at $\xi=0$ to $\phi_{\nu\alpha}^\xx(\rr)=\phi_{\nu\alpha}(\rr)$ at $\xi=1$ remains arbitrary.

In contrast to the other three conventional routes to thermodynamics [see Eqs.\ \eqref{1.1}, \eqref{1.2}, and \eqref{1.5}], which only need the pair correlation functions of the solvent system, the chemical-potential route requires the solute-solvent pair correlation functions for every value of the coupling parameter $\xi$. This implies that, even in the one-component case, this fourth route makes use of the pair correlation function of a binary mixture, albeit  one of the species (the solute) is present with a vanishing concentration. This inherent multicomponent character of the chemical-potential route can in principle hamper its practical implementation.

As is well known, thermodynamic quantities obtained from a common \emph{approximate} pair correlation function via independent routes are not necessarily consistent. In particular, the thermodynamic relations \eqref{1.11}--\eqref{1.13} can be violated when the left-hand sides are evaluated from Eq.\ \eqref{2.22} and the right-hand sides are evaluated from Eqs.\ \eqref{1.2}, \eqref{1.1}, and \eqref{1.5}, respectively. Moreover, the chemical-potential route introduces extra sources of possible thermodynamic inconsistencies. First, in the case of a mixture, the symmetry condition \eqref{1.10} may not be fulfilled. This is a consequence of the fact that, in contrast to the energy, virial, and compressibility routes, where global quantities are obtained, the chemical-potential route provides a quantity for each separate component of the mixture. On the other hand, Eq.\ \eqref{1.10} is trivially satisfied by one-component systems. A second source of inconsistency is much subtler. As said before, the chemical-potential route implies to load a new particle into the system by means of a coupling parameter $\xi$ and it is not guaranteed that the final result will be independent of the protocol followed in the loading process, as it should be. {}From that point of view, given an approximate theory, this route is expected to provide a whole class of results, rather than a unique one, even for one-component systems. {This protocol-related thermodynamic inconsistency is indeed observed in the PY solution of Baxter's sticky-hard-sphere model \cite{B68}, as will be extensively analyzed in a forthcoming paper \cite{RS13}.}

The general scheme has been particularized to the prototype HS fluid mixture. In this case, since the interaction potential $\phi_{\alpha\gamma}(r)$ depends on a single parameter $\sigma_{\alpha\gamma}$, it is possible to eliminate the coupling parameter $\xi$ in favor of the solute-solvent interaction range $\sigma_{0\alpha}$, thus avoiding the protocol problem alluded to above. The result, given by Eq.\ \eqref{4.10}, is valid for any HS mixture, both additive and nonadditive. Further progress can be made if $\sigma_{\alpha\gamma}\geq \frac{1}{2}(\sigma_\alpha+\sigma_\gamma)$, i.e., if the HS mixture is either additive or has a positive nonadditivity. In such a case, the evaluation of the contribution to the chemical potential when $0\leq\sigma_{0\alpha}\leq \frac{1}{2}\sigma_\alpha$ amounts to the trivial computation of the volume accessible to a point particle in a sea of non-overlapping solvent spheres.

As a practical implementation of the chemical-potential route, we have considered the SPT and PY approximations for three-dimensional additive HS mixtures.
The SPT result is consistent with the symmetry condition \eqref{1.10}, while the PY result is not. This represents a neat example showing that the internal consistency condition \eqref{1.10} is not guaranteed by an approximate RDF. Moreover, the SPT chemical potential is consistent with the SPT virial equation of state (which coincides with the PY compressibility equation of state), but the PY chemical potential yields an equation of state that differs from the PY virial equation. In fact, the former turns out to be more accurate than the latter, as comparison with computer simulations reveals. Interestingly, an interpolation  between the PY compressibility and chemical-potential routes provides slightly better results than the conventional interpolation  between the PY compressibility and virial routes giving rise to the BMCSL equation of state.}

\acknowledgments
{A.S. acknowledges support from the Spanish government through Grant No.\ FIS2010-16587 and  the Junta de Extremadura (Spain) through Grant No.\ GR10158, partially financed by Fondo Europeo de Desarrollo Regional (FEDER) funds.
The work of {R.D.R.} has been supported by the Consejo Nacional de Investigaciones Cient\'ificas y
T\'ecnicas (CONICET, Argentina) through Grant No.\
PIP 112-200801-01474.}

\bibliographystyle{apsrev}

\bibliography{D:/Dropbox/Public/bib_files/liquid}

\begin{thebibliography}{24}
\bibitem[{not()}]{note_13_03}
\bibinfo{note}{Notice that, since $A$ is obtained from $\Psi_i$ by integration
  over a certain thermodynamic variable, an unknown function remains in $A$
  under the form of an integration constant. If the integration is carried out
  over density, the unknown function can be determined from the ideal-gas free
  energy.}
\end{thebibliography}

\end{document}